# The Search for Primordial Black Holes Using Very Short Gamma Ray Bursts

D.B. Cline[*], C. Matthey and S. Otwinowski

*University of California Los Angeles*
*Department of Physics and Astronomy*
*Box 951447, Los Angeles, California 90095-1547 USA*

B. Czerny, A. Janiuk

*Copernicus Astronomical Center*
*ul. Bartycka 18, 00-716 Warsaw, Poland*

**Abstract.** We show the locations of the SWIFT short hard bursts (SHB) with afterglows on the galactic map and compare with the VSB BATSE events. As we have pointed out before, there is an excess of events in the galactic map of BATSE VSB events. We not that none of VSB SWIFT era events fall into this cluster. More SWIFT events are needed to check this claim. We also report a new study with KONUS data of the VSB sample with an average energy above 90 keV showing a clear excess of events below 100 ms duration ($T_{90}$) that have large mean energy protons. We suggest that VSB themselves consist of two subclasses: a fraction of events have peculiar distribution properties and have no detectable counterparts, as might be expected for exotic sources such as primordial black holes. We show how GLAST could add key new information to the study of VSB bursts and could help test the black hole concept.

## INTRODUCTION

Gamma Ray Bursts were early recognized as consisting of two separate populations: Long Bursts and Short Burst [1]. Long bursts ($T_{90}$ > 2 s) are known to originate at cosmological distances with many identified counterparts, and they are widely believed to be related to the collapse of massive stars [2-4]. Short Gamma Ray bursts properties were also extensively studied [5-9] but their nature is less clear. Over the years there have been many concepts put forward for the origin of SHB, including the mechanism of primordial black holes (PBH) evaporation [10-15]. One of the recent SHB (GRB 060313) shows rather peculiar time behavior, difficult to reconcile with expectations from the binary merger; see [16, 17, 19-23, 25-28].

In this article we discuss some new properties of very short bursts (VSB; $T_{90} \leq 100$ ms) from BATSE, KONUS and SWIFT. In earlier publications [25-27, 29-31] we have found some unusual properties of VSB events, including very hard photon spectra compared to longer duration GRBs, a significant angular asymmetry on the Galactic plane and a ‹V/V$_{max}$› value consistent with 0.5. These properties, and the new results presented in this paper suggest that a part of VSB events can originate from PBH decay [11, 26, 29, 31-35].

## COMPARISON WITH CURRENT AND FUTURE SWIFT DATA

The angular distribution in Galactic coordinates is shown in Fig. 1 for all, 51, VSBs events ($T_{90} \leq 100$ ms) from BATSE – full circles. We divided the sky into eight equal regions. In one of the regions (roughly in the direction of

---

the Galactic anticenter) we observe 20 events, which is much higher then the expected average of 51/8. The probability of finding 20 or more events (from the total number of 51) in the region of _ of the whole sky is 0.00007. This result argues for an explanation other than the statistical fluctuation. The background in the direction of the Galactic center is $12,500 \pm 1000$ counts s$^{-1}$, while the mean level of the background outside the excess region is $13,800 \pm 1300$ counts s$^{-1}$, and the total number of short bursts (SB) (100 ms < $T_{90}$ < 2 s) in this region is slightly lower than the expected average (but within the expected error). Therefore, the background anisotropy cannot be responsible for the observed angular distribution of the VSBs across the sky.

We have now put in Fig. 1 also the short hard bursts (SHB) from Swift, with afterglows (open triangles) and without afterglows (open squares). There are four VSB among them, two with detected afterglows and two without an afterglow.

The two VSB events with afterglows from SWIFT are not in the excess region of events from BATSE (Fig. 1). This may imply that these events come from different source type. The third one, without an afterglow, is located not far from the Galactic center, and the one is actually quite close to the anticenter region, although not quite within it. Among all 10 events SHB from SWIFT none is located exactly within the anticenter region. Such a distribution is still consistent with random distribution across the sky. On the other hand, BATSE statistics would suggest that 1 out of 3 VSB should be located in the anticenter region. Since there is none, we suggest that SWIFT instrument is likely to be detecting other, more soft bursts, so the relative increase in the fraction of SHB with respect to the whole population, and in fraction of VSB with respect to SHB, does not translate into an increase in the number of bursts in the anticenter region, and there is no contradiction between the current Swift SHB data and the VSB data from BATSE.

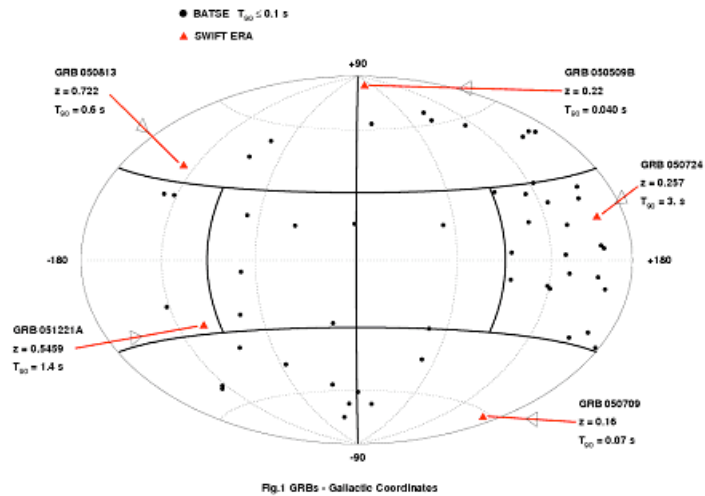

**FIGURE 1.** The map of the sky in Galactic coordinates. Black dots mark the VSB from BATSE, triangles mark new SWIFT/HETE events with afterglows and squares mark VSB from SWIFT without afterglows. One event on the plot has a $T_{90}$ of 3 sec.

The KONUS data also contains a number of SHB events with extended X ray emission – one of those events was in the VSB category (Table 2). For now it is possible that the excess events in Fig. 1 form a different population than most of the newly discovered SWIFT events: they are possibly less likely to develop strong X-ray afterglows and their emission is likely to extend to higher energies. The lack of extension of the SWIFT detectors toward high energies, the small number of these events and no localization information for the KONUS events prevent us from testing the hypothesis directly. However, we provide some additional arguments in favor of the existence of two separate classes of VSB in the next sections.

# EXCESS OF VSB EVENTS IN KONUS DATA FOR HIGH ⟨Eγ⟩

We have recently published a paper [31] comparing the BATSE and KONUS data. This clearly shows that the KONUS events with $T_{90} < 100$ ms have a higher energy photon component. All VSBs show an appreciable number of photons above 1 MeV energy. All events also show gamma rays above 5 MeV. This follows the trend observed in the BATSE data that the VSB events have hard energy spectrum. In the paper [31] we compared the mean energy of SBs and VSBs for KONUS events. We observe there, that in the MeV region, the spectra of VSBs are significantly harder than the spectra of SBs. The spectra start to be flatter above 3 MeV, and the effect in the case of VSBs is again stronger.

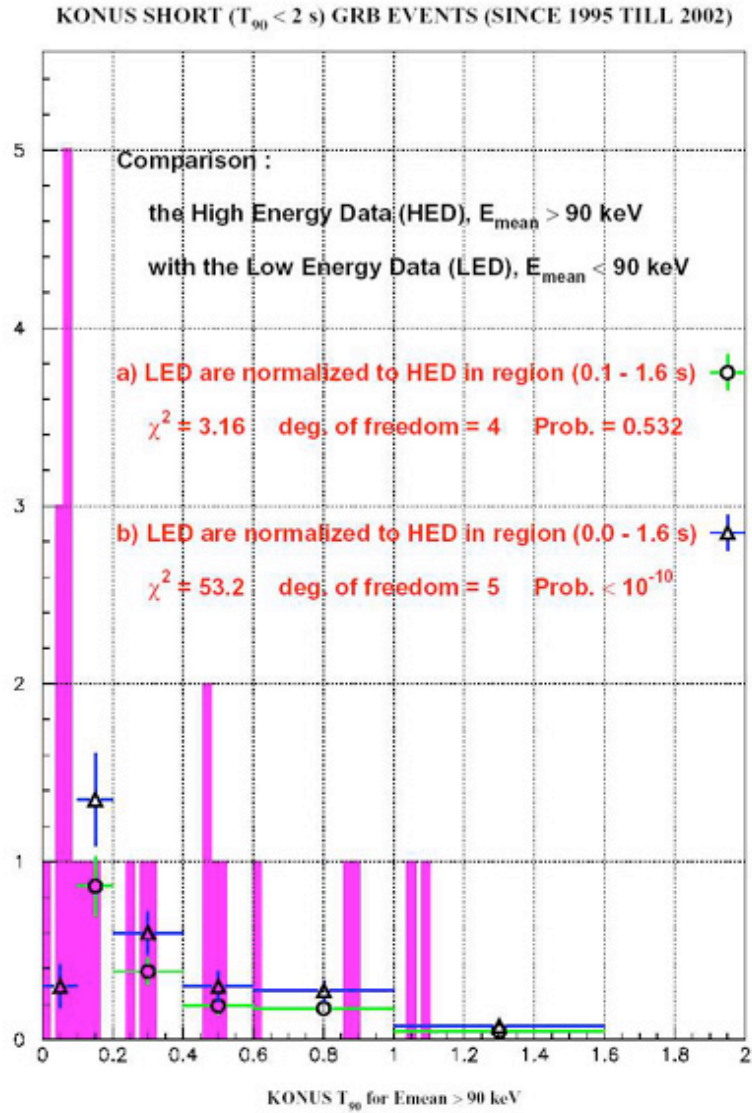

To follow up we study all the SHB events in KONUS data and select burst with the mean energy ⟨E_⟩ > 90 keV. We constructed a histogram of burst numbers as a function of their duration. For comparison, we show similar histogram made for SHB with mean energies ⟨E_⟩ < 90 keV, and scale it down to the distribution of harder bursts. Comparing these two distributions we see very strong clumping of hard bursts at very short durations. In the histogram time interval of 0-0.1 for T90 we expect one event from softer bursts distribution and found ten hard events, which is extremely unlikely---indication again of some new physics origin of the bulk of the VSB data.

The KONUS detector has a significantly larger energy acceptance out to tens of MeV. The BATSE detector has a smaller NaI absorber and is not sensitive to 1-10 MeV photons, but measures event coordinates, which KONUS not measures.

We have recently published a paper [31] comparing the BATSE and KONUS data. This clearly shows that the KONUS events with $T_{90} < 100$ ms have a higher energy photon component. All VSBs show an appreciable number of photons above 1 MeV energy. All events also show gamma rays above 5 MeV. This follows the trend observed in the BATSE data that the VSB events have hard energy spectrum. In the paper [31] we compared the mean energy of SBs and VSBs for KONUS events. We observe there, that in the MeV region, the spectra of VSBs are significantly harder than the spectra of SBs. The spectra start to be flatter above 3 MeV, and the effect in the case of VSBs is again stronger.

To follow up we study all the SHB events in KONUS data and select burst with the mean energy ‹E_› > 90 keV. We constructed a histogram of burst numbers as a function of their duration (see Fig. 2b). For comparison, we show similar histogram made for SHB with mean energies ‹E_› < 90 keV, and scale it down to the distribution of harder bursts. Comparing these two distributions we see very strong clumping of hard bursts at very short durations. In the histogram time interval of 0-0.1 for T90 we expect one event from softer bursts distribution and found ten hard events, which is extremely unlikely---indication again of some new physics origin of the bulk of the VSB data.

The KONUS data also included a number of SHB events with extended X ray emission (see results at the web site Http://www.physics.ucla.edu/hep/grb/table2.jpg). It is thus possible that the excess events in Fig. 1 do not have a strong X ray afterglow unlike the detector SWIFT events.

Note – More VSB in SWIFT then expected but none in the excess region – this is **improbable** if **all VSB have counterparts**.

In the previous analysis of BATSE data using the TTE sample [25] we showed that the composite of the VSB time profile made of 12 bursts has a different time profile than other SB. We now construct the new composite profile of the VSB from the whole BATSE sample (see Fig. 3). We fit the profile as an exponential rise and a decay law of (1 – t/td)n, as in [9]. The best fit value of the index n equal to 1.00 is lower than the median value of 1.30 for the distribution of SB [9]. This results supports the view that VSB profiles are systematically different from SB profiles.

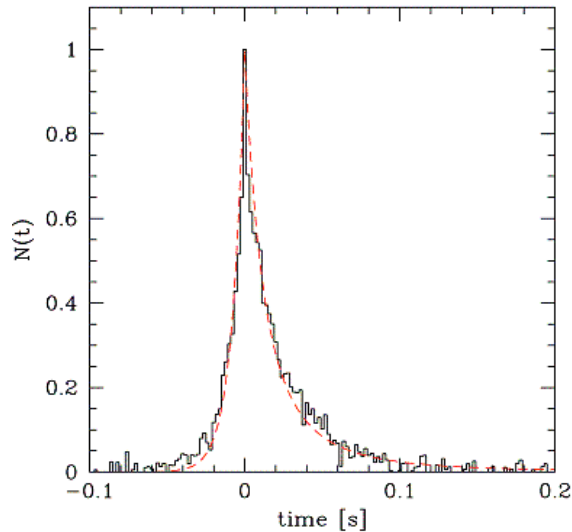

Fig. 3. Composite profile made out of background-subtracted and normalized BATSE VSB bursts together with analytical fit to the profile (exponential rise with $t_r$=0.041 s, decay with $t_d$=0.032 s, and power law index n = 1.00).

## CONCLUSION

In this note we have documented two new aspects of the VSBs:

(a) The two VSB Swift/HETE events are not located in the excess region observed in the BATSE data. It is not possible to measure the high energy part of the energy spectrum so we cannot test whether these events are in the same class of the BATSE VSB or not.

(b) A new study of KONUS data indicates that VSB events are much harder than the rest of the SHB events, strongly suggesting a new physics origin of these events. The results of our enhancement below T90 of 100 ms is confirmed by other types of analysis of BATSE data discussed here. This likely indicates some new source of these events such as primordial black hole evaporation in the Galaxy, not far from the Solar System.

(c) New studies with GLAST could sharpen the case for PHBH origin.